\documentclass{moriond}

\def\Journal#1#2#3#4{{#1} {\bf #2}, #3 (#4)}

\def\PLB{{\em Phys. Lett.}  B}

\def\PRD{{\em Phys. Rev.} D}

\def\apss{\em Aph. Sp. Sci.}
\def\jcap{\em  J. Cosmol. Astropart. Phys.}
\def\aap{\em Astron. Aph.}
\def\mnras{\em MNRAS}
\def\azh{\em Astron. Zh.}
\def\apj{\em Aph. J.}

\def\jcap{\em J. Cosmol. Astropart. Phys.}

\def\be{\begin{equation}}
\def\ee{\end{equation}}
\def\bea{\begin{eqnarray}}
\def\eea{\end{eqnarray}}

\usepackage{epic}
\usepackage{epsfig}
\usepackage{setspace}

\newcommand{\Photo}{\vspace{35mm}}

\begin{document}
\vspace*{4cm}
\title{DISTANT FOREGROUND AND THE HUBBLE CONSTANT TENSION}

\author{ V.N. YERSHOV }

\address{Moniteye U.K., 12 Ogle Street, London, W1W 6HU}

\vspace{-5.0cm}
\hspace{-1cm}
\begin{minipage}{15cm}
\begin{spacing}{0.8} 
{\scriptsize 
Proceedings of the 56$^{th}$ Rencontres de Moriond - 2022 "Cosmology", La Thuile, Italy, Jan 23-30 2022 \\
E.A.J. Dumarchez \& J. Tr\^{a}n Thanh V\^{a}n (eds.), ARISF Publ., 2022, 237-240 \\
}
\end{spacing}
\end{minipage} 
\vspace{3.5cm}

\maketitle\abstracts{
It is possible to explain the discrepancy (tension) between the local measurement of the cosmological 
parameter $H_0$ (the Hubble constant) and its value derived from the {\it Planck}-mission 
measurements of the Cosmic Microwave Background (CMB) by considering contamination 
of the CMB by emission from some medium surrounding distant extragalactic sources
(a distant foreground), such as  
extremely cold coarse-grain (grey) dust.  
As any other foreground, it would alter the CMB power spectrum 
and contribute to the dispersion of CMB temperature fluctuations. 
By generating random samples of CMB with different dispersions,  
we have checked that the increased dispersion leads to a smaller estimated value of $H_0$, 
the rest of the cosmological model parameters remaining fixed. This might explain the 
reduced value of the {\it Planck}-derived parameter $H_0$ with respect to the local measurements.  
The cold grey dust for some time has been suspected to populate intergalactic 
space and it is known to be almost undetectable, except for the effect 
of dimming remote extragalactic sources.}

\setcounter{page}{237}

\section{Introduction}
The importance of the issue with the Hubble constant as measured by 
two different methods (the $H_0$ tension) can be appreciated from 
recent comprehensive reviews on the subject
\cite{divalentino21,shah21} and by the fact 
of special international conferences discussing exclusively this 
particular issue \footnote{\texttt{https://www.eso.org/sci/meetings/2020/H0.html}}. 

The {\it Planck} space observatory \cite{aghanim20} revealed a 
statistically significant discrepancy between the  
cosmological parameter $H_0$ as calculated 
within the standard  
cosmological model by using the Cosmic Microwave Background (CMB) power spectrum,
$H_0=(67.37\pm0.54)$
\,km\,s$^{-1}$\,Mpc$^{-1}$,
and the values of this parameter obtained by using other methods -- 
mostly from direct local measurements \cite{riess20}. 
One of these local measurements is based on optical and infrared (IR)
observations of variable Cepheid stars, 
with the recent calculation of $H_0$ based on this method \cite{riess18} being 
$H_0=(73.48\pm1.66)$\,km\,s$^{-1}$\,Mpc$^{-1}$.
Both local and {\it Planck}-derived estimates of $H_0$ have passed a number 
of rigorous tests by considering many possible sources of systematic errors
\cite{efstathiou13,planck17,follin18},
but the discrepancy still remains.

Discussing the possible origin of this discrepancy, most authors and
reviewers  focus primarily on 
observational biases related to the method of standard-candles,  Cepheids and type-Ia 
supernovae (SN), and on proposals going far beyond the standard cosmological and 
particle physics models. By contrast, possible biases intrinsic to the 
CMB are passed by almost 
without further thought.    

\begin{figure}
\vspace{-0.5cm}
\hspace{-0.3cm}
\epsfig{figure=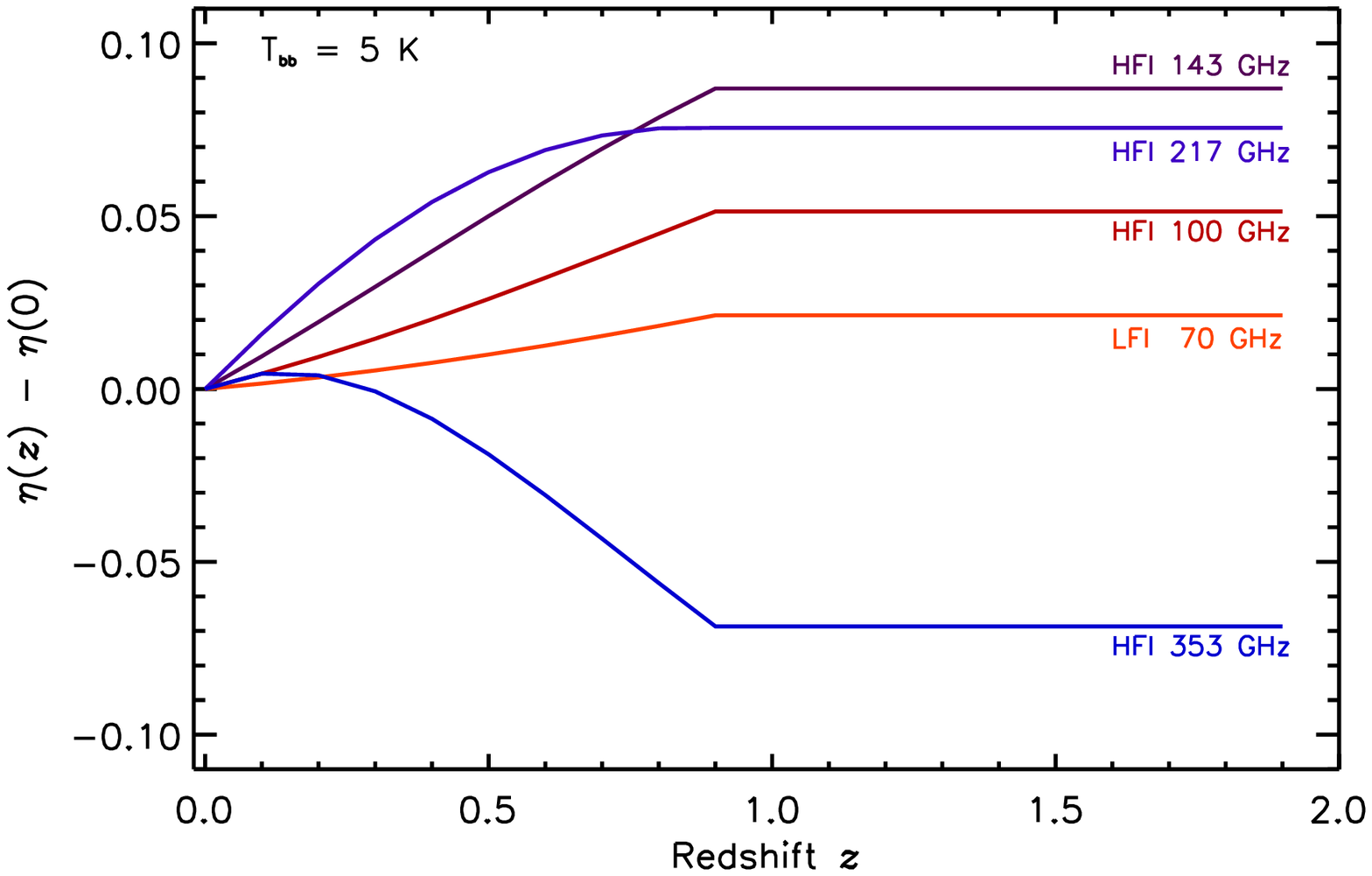,width=7.5cm}
\hspace{-0.1cm}
\epsfig{figure=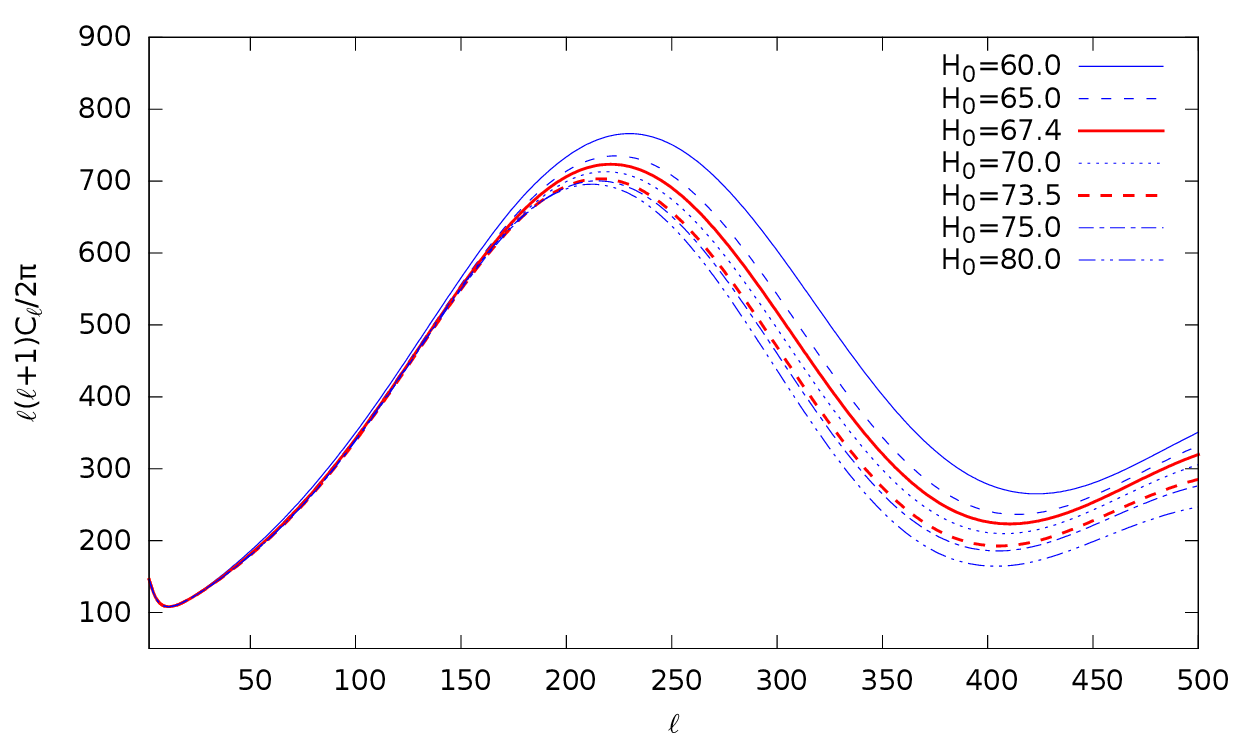,width=8.5cm}
\vspace{-0.3cm}
\caption{{\small {\it Left:} Redshift dependences of the blackbody radiation energy
fraction $\eta(z)$ as observed in five {\it Planck} 
frequency bands for $T_{\rm bb}=5$\,K; the constant shifts of the curves
with respect to each other have been normalised at $z=0$ by 
subtracting from them their individual values $\eta(0)$;
{\it Right:}~CMB~power spectra (in the standard normalised presentation) generated by using 
the code for anisotropies in the microwave background (CAMB tool) 
for seven different values of $H_0$.}} 
\label{fig:bb_slopes}
\end{figure}

\section{Distant foreground}
Various authors have reported that the contaminating emission from a medium 
around distant extragalactic sources should affect correlations between CMB and 
extragalactic cosmic structures traced by bright transient sources,
like supernovae (SNe) or gamma-ray bursts.   
The signature of the distant foreground in the CMB, 
based on the WMAP and {\it Planck}-mission results and traced by SNe was previously 
reported by the author \cite{yershov12,yershov14}, 
who found a correlation 
between the SN redshifts, $z_{\rm SN}$,  and CMB temperature fluctuations 
at the SNe locations, $T_{\rm SN}$. 

By using different {\it Planck} frequency bands and computing 
the fractions $\eta_\nu(z)$  of blackbody radiation energy as observed in each 
frequency band $\nu$ for different redshifts $z$ one can estimate the regression
line slopes 
for these bands \cite{yershov20}. The functions $\eta_\nu(z)$ are shown
in Fig.\,\ref{fig:bb_slopes}, as calculated for the blackbody temperature 
$T_{\rm bb} = 5$\,K and five of the {\it Planck} frequency bands 
$\nu= \{70, 100, 143, 217, 353\}$\,GHz. Note that the temperature of the 
medium in thermal equilibrium with the CMB must exceed or be equal to 5\,K for the redshifts 
$z > 0.835$, so for these redshifts the functions $\eta_\nu(z)$ appear 
horizontal.

We can calculate slopes $\xi^c_\nu$ of these functions for different $T_{\rm bb}$ 
and compare them with the observed slopes $\xi^o_\nu$ corresponding to 
different {\it Planck} frequency bands $\nu= \{70, 100, 143, 217, 353\}$\,GHz which,
according to our previous work \cite{yershov14}, are the following: $\xi^o_{70}=0.96\pm 0.63$, $\xi^o_{143}=1.09\pm 0.31$,
$\xi^o_{217}=0.61 \pm 0.36$ and $\xi^o_{353}= -0.99\pm 0.48$. 
The slope for the 100\,GHz-band 
was used as a reference for normalisation, so $\xi^o_{100}=1.00\pm 0.39$.

The calculated slopes $\xi^c_\nu$ of these functions averaged between the 
temperatures $T_{\rm bb} = \{3, 4, 5, 6\}$\,K for each 
of the {\it Planck} frequency bands $\nu= \{70, 143, 217, 353\}$\,GHz 
are
%
$\xi^c_{70}=0.45\pm 0.03$, $\xi^c_{143}=1.63\pm 0.17$,
$\xi^c_{217}=0.73 \pm 0.74$ and $\xi^c_{353}= -1.48\pm 0.43$
(again, the slope for the 100\,GHz-band 
was used as a reference).
They are within the 1-$\sigma$ tolerance interval with respect
to the above experimental values $\xi^o_\nu$, which indicates that the 
temperature of the CMB-contaminating ingredient of the 
intergalactic medium is very low, likely to be between 3\,K and 6\,K. 
This can give clues as to the nature of the medium, which can be 
coarse-grain (grey) dust, and which for some time has been suspected to 
populate intergalactic space  
\cite{eigenson49,zwicky57,gonzalez98,alton01}.

This ``grey'' dust
leaves little or no imprint on the spectral energy distribution of background sources.  
However, it creates the long-known excess of radiation from some extragalactic objects in the 
far IR at $\lambda \approx 500\,\mu$m, which extends up to centimetre wavelengths and
can interfere with the CMB radiation. Such a  $500\,\mu$m radiation excess 
was confirmed and measured by the 
{\it Herschel} space observatory \cite{galliano11}. 
In the 1990s, this excess was interpreted as an elevated spatial mass 
density of cold dust \cite{reach95} with temperatures of 4 to 7\,K. 
Here we confirm this interpretation from a completely different 
point of view.

\section{CMB distrortion}
The angular sizes of the observed regions with the 500\,$\mu$m emission \cite{galliano11,lisenfeld02}
range from 0.02$^\circ$  to 0.5$^\circ$.
So this emission would effectively distort the CMB power spectrum at the 
multipole moments $\ell \approx 360$ and higher, which would change
the estimated parameter $H_0$.  
In order to quantify these changes we have used 
the code for anisotropies in the microwave background \cite{lewis13} (CAMB)  
%
which allows the extraction of different cosmological parameters 
from theoretical CMB power spectra generated by the same code. 
The calculated changes  
 are shown on the right panel of Fig.\,\ref{fig:bb_slopes}
for the first trough of the CMB power spectrum,
where its effect  on the calculated parameter $H_0$
is quite strong. 

In this code, the coefficients  $C_\ell$ of the CMB power spectrum are calculated as 
sums of the integrals $a_{\ell m}$, $|m| \le \ell$, which include
temperature fluctuations $\Delta T(x,\phi)$ over the celestial sphere, where
$x\in[-1,1]$ is the cosine of the latitude and $\phi\in[0,2\pi]$ is the longitude.
Conversely, the functions $\Delta T(x,\phi)$ are calculated by summing up the integrals
$a_{\ell m}$.  
For a given CMB power spectrum, we have calculated a set of corresponding values of 
$\Delta T(x,\phi)$
by using random $a_{\ell m}$ for 
$\ell=0,1,\dots,\ell_{\tt max}$ with the 
restriction $\ell_{\tt max}=500$.
We have taken five equal-spaced values of $H_0$, namely, 
60, 65, 70, 75 and 80 [km s$^{-1}$ Mpc$^{-1}$]  
plus the values 67.4 [km s$^{-1}$ Mpc$^{-1}$] (solid red curve on the right 
panel of Fig.\,\ref{fig:bb_slopes}) 
and 73.5 [km s$^{-1}$ Mpc$^{-1}$]
(dashed red curve on the same panel) 
corresponding, respectively, 
to the {\it Planck} result and to the local measurements of $H_0$.

Additionally, for checking the consistency of our calculations we 
have taken a few sets of normally distributed randomised values of
$a_{\ell m}^i$, $i=1,2,\dots,5$, so that for each of the selected 
values of $H_0$, we have obtained five samples of $a_{\ell m}^{H_0,i}$ and, 
correspondingly,  five samples of values $\Delta T^{H_0,i}$.
For each of them, we have calculated the average of the CMB temperature 
fluctuations $\overline{\Delta T}$ and its standard deviation $\sigma_{T}$.
Here we are mainly interested in the way the values $\sigma_T$ change 
when the parameter $H_0$ is varied.  
For each of these generated sequences, the trend of the 
calculated values $\sigma_{T}$ was practically the same.
Namely, when the dispersion of the CMB temperature fluctuations increases, the 
value of the estimated $H_0$ decreases, the difference between the two 
discussed $H_0$ values 73.5 and 67.4  [km s$^{-1}$ Mpc$^{-1}$] being related to
$\Delta\sigma_{T}=-0.60\pm 0.04\,\mu$K.
This value is the measure of CMB contamination by 
photons from the medium surrounding remote clumps of matter,
and it can thus be used for estimating the amount of cold 
coarse-grain dust in the intergalactic medium. 

\section{Discussion}

Between 2012 and 2019, new studies have appeared demonstrating that the dimming 
of the type-Ia supernovae was different in different directions 
\cite{cai12,cardenas13,bernal17,colin19,zhao19,rameez19,luongo21}.
Most of the authors of these 
studies interpret their results in terms of anisotropic acceleration of the Universe. 

However, anisotropy of acceleration violates the main cosmological principle. Therefore,
the grey-dust interpretation of the type-Ia supernovae anisotropic dimming becomes preferable:
it would be much more logical to assume non-uniform distribution of dust 
rather than the Universe having different properties in different 
directions.

What might be the origin of this coarse-grain dust?
Apart from the initial formation of dust particles within galaxies with 
their further transport into the intergalactic space, 
dust formation can also occur  
directly in the intergalactic medium \cite{fabian94}. 
The observational evidence for cold molecular clouds at the cooling flows
in galaxy clusters and the presence of dust in these regions is widespread
\cite{russell17}. 
Besides this detectable dust clouds, 
there exists yet another possibility of almost undetectable  
coarse-grain dust existing in the intergalactic space which, 
according to recent studies, should be seriously taken into 
consideration. Namely, these macroscopic dust particles can be formed 
by the process of hydrogen solidification under sufficiently low 
temperatures, e.g., when the CMB temperature gets below 10\,K at $z<2$,    
which was proposed in 1968 by F. Hoyle \cite{hoyle68} and 
further discussed in the 1990s and early 2000s
 \cite{pfenniger94,wardle99,pfenniger04}. 
In the last decade, it was shown by experimental physicists that H$_2$ ice 
became stable in vacuum and do not sublime if it 
contains impurities \cite{schaefer07,walker13,kettwich15} 
transported from galaxies into the intergalactic medium.

My conclusion is that the mechanism of contamination of CMB radiation 
by some distant foreground emission from cold dust can explain the discrepancy 
between the local measurements of $H_0$ and the {\it Planck}-derived value, 
without invoking unnecessary assumptions that violate the basic cosmological 
principles or break the standard cosmological and particle physics models.

\end{document}